# Translating AI into scientific impact: Field context, career position, and institutional capability in AI-enabled research

Zhiyong Tan, Hongkan Chen [*], Yi Bu [*]

*Department of Information Management, Peking University, Beijing 100871, China*

**Correspondence concerning this article should be addressed to Hongkan Chen (chenhongkan@pku.edu.cn) and Yi Bu (buyi@pku.edu.cn)**

**Abstract:** Artificial intelligence (AI) is increasingly embedded in scientific research, but its scientific value is unlikely to be distributed evenly. This study examines how AI knowledge integration is associated with scientific impact and asks who benefits from AI-related knowledge in science. Using large-scale bibliographic data, we measure AI integration through references to papers in the OpenAlex Artificial intelligence subfield and link it to five-year citation impact. The results show that AI references are generally associated with higher citation impact, but the returns vary substantially across scientific fields. Career stage also matters: senior scholars benefit more from the extensive margin of AI referencing, whereas junior scholars benefit more from intensive AI referencing and tend to cite newer and higher-impact AI papers. At the institutional level, returns are non-monotonic: institutions with intermediate AI capability achieve the largest proportional gains, while leading AI institutions are more deeply embedded in AI-centered knowledge spaces, attract more AI-related audiences, and more often become substitute citation gateways to cited AI sources. These findings suggest that the value of AI knowledge depends not only on technical capability, but also on translational capacity: the ability to make AI knowledge meaningful, legitimate, and useful across scientific communities.

## 1. Introduction

Artificial Intelligence (AI) is rapidly becoming an important knowledge input into scientific research (Liu et al., 2025; Wang et al., 2023); it has become an object of strategic investment and organizational redesign in science. Governments and universities are investing heavily in AI expertise and computing infrastructure, as reflected in national AI initiatives in China and the United States (Ministry of Education of the People's Republic of China 2018; State Council of the People's

Republic of China 2017; National Science Foundation 2023, 2024; U.S. Congress 2021). These investments reflect a widespread expectation about the transformative potential of AI-enabled research, and consistent with this expectation, prior studies associate AI-enabled research with changes in scientific productivity, novelty, and impact (Bianchini et al., 2026, 2022; Gao and Wang, 2024; Liu et al., 2025; Wang et al., 2023). Yet the growing availability of AI does not imply that researchers and organizations will obtain equal returns from it.

Management research has long emphasized that the value generated by a technology depends on complementary assets, skills, and organizational arrangements rather than on the technology alone (Milgrom and Roberts, 1990; Teece, 1986). We extend this insight from access to AI technology to the integration of AI knowledge in scientific research. The conversion of AI knowledge into scientific contributions and impacts are similarly embedded in multiple evaluative and organizational contexts. Scientific contributions and impacts are evaluated within different fields, produced by researchers occupying different positions, and embedded in institutions with different knowledge bases and audience ties. Research on digital inequality similarly distinguishes access to a technology from differences in how it is used and the outcomes obtained from that use (DiMaggio and Hargittai, 2001; van Deursen and Helsper, 2015). This raises our central question: Who benefits from AI knowledge, and how do field, career, and institutional contexts shape its translation into scientific impact?

We develop a multilevel framework for understanding these unequal returns. At the field level, scientific communities differ in their methodological traditions, problem structures, and criteria of legitimate contribution (Suchman, 1995; Whitley, 2000). At the individual level, career stages shapes researchers' resources, visibility, and the meanings attached to their engagement with an emerging technology (Zuckerman and Merton, 1972). At the institutional level, AI capability affects the ability affects the ability to identify, understand, and incorporate specialized knowledge. These three contexts jointly shape whether engagement with AI knowledge is regarded as relevant, credible, and consequential.

Overall, we argue that AI-enabled science is not merely about who knows AI best, but also who can make it matter. The rise of AI in research therefore creates not only new opportunities, but also new forms of stratification, brokerage, and translation in the production of scientific knowledge.

This study makes two contributions. First, it shifts research on AI in science from adoption and average effects to unequal returns from the same technological knowledge input. Second, it develops a multilevel account of how disciplinary context, career position, and institutional capability shape the translation of AI knowledge into scientific impact. In doing so, it offers new insights into how AI is reshaping the organization of science.

## 2. Theoretical background and hypotheses development

### *2.1. AI knowledge integration and scientific impact*

Scientific research advances through the recombination, adaptation, and extension of existing knowledge. Since Schumpeter's classic view of innovation as new combinations (Schumpeter, 1983), a large body of research has emphasized that scientific and technological progress often emerges when actors recombine knowledge components across domains (Fleming, 2001; Uzzi et al., 2013). In this sense, the value of a knowledge input depends not only on its intrinsic quality, but also on whether it can be mobilized to generate new problems, methods and interpretations. From the perspective of structural hole theory, actors who bridge otherwise disconnected knowledge domains can access diverse and non-redundant information, thereby gaining advantages in generating novel combinations and achieving higher impact (Burt, 1992). In scientific research, referencing knowledge from different domains may position a paper in a brokerage role, enabling it to connect disparate intellectual communities and enhance its potential for recognition and influence.

AI knowledge is increasingly important in this process because it has features of a general-purpose input (Agrawal et al., 2024; Bianchini et al., 2022). General-purpose technologies are characterized by wide applicability, continuous improvement, and complementarities with other domains of activity (Bresnahan and Trajtenberg, 1995). AI methods and models can be applied to a wide range of scientific tasks, including prediction, classification, image recognition, text analysis, and so on.

Papers that cite AI-related literature may therefore gain scientific impact through two related but distinct margins. At the **extensive margin**, the presence of any AI-related references indicates that a paper connected its problem, data, or method to an expanding body of AI knowledge. This may allow them to formulate new problems or revisit existing problems with new methodological tools. At the **intensive margin**, the share of AI-related references captures the depth of this engagement. A greater reliance on AI knowledge may reflect stronger absorptive capacity and deeper methodological integration, which can increase a paper's ability to recombine domain knowledge with computational techniques (Cohen and Levinthal, 1990). The paper may also draw on a broader set of AI knowledge components, expanding the opportunities for recombining knowledge in potentially valuable ways. However, this effect may not be linear: excessive reliance on AI-related knowledge may reduce the balance between conventional domain grounding and atypical recombination, thereby weakening the broader scientific appeal of the paper (Uzzi et al., 2013).

The starting point of our argument is modest: on average, papers that connect themselves to AI knowledge are expected to receive greater attention. We thus propose:

**Hypothesis 1a.** Papers that cite AI-related references have higher citation impact than otherwise similar papers that do not cite AI-related references.

**Hypothesis 1b.** The share of AI-related references in a paper's reference list is positively associated with its citation impact.

**Hypothesis 1c.** The positive association between the share of AI-related references and citation impact is subject to diminishing marginal returns.

### *2.2. Unequal returns to AI knowledge: A multilevel framework*

The value of AI knowledge is unlikely to be determined by AI references alone, because once AI knowledge enters a paper, its value must be recognized and evaluated again. This process is shaped by multiple layers of context. At the paper level, the disciplinary field of the focal paper determines whether AI-based knowledge is viewed as legitimate and useful. At the individual level, the author's career stage shapes whether AI references are interpreted as a signal of renewal, competence, or frontier engagement. At the institutional level, AI capability affects not only the ability to absorb AI knowledge, but also the ability to transform and reuse it. We therefore develop a multilevel framework of unequal returns to AI knowledge integration as illustrated in **Figure 1**.

**Figure 1 Multilevel framework for the unequal returns to AI knowledge**

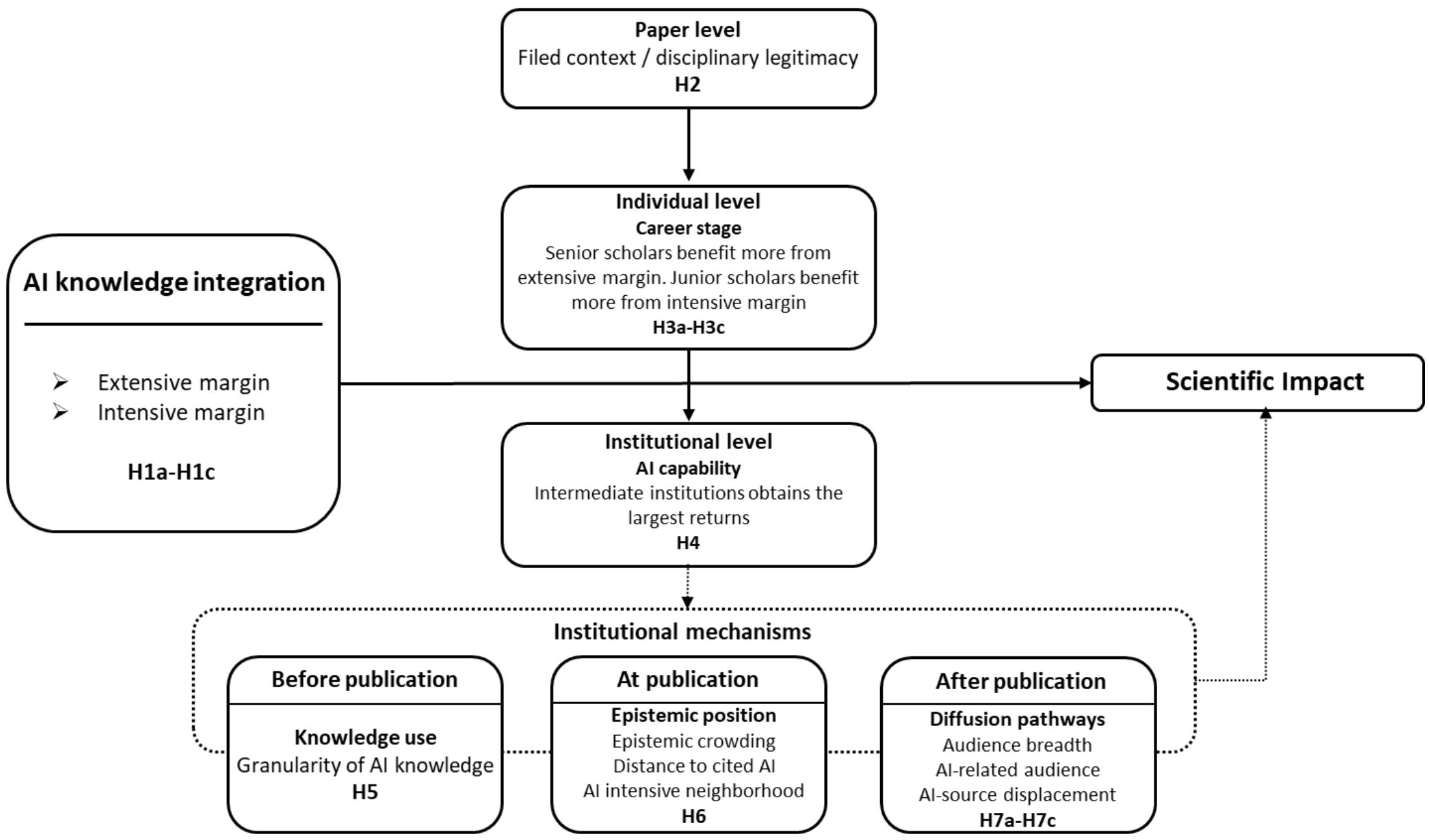

*Notes.* The figure illustrates our hypothesized framework for the unequal returns to AI knowledge.

### 2.2.1. *Paper-level disciplinary context: field legitimacy*

Although AI may serve as a general-purpose input, scientific fields differ in how they evaluate external knowledge. Science is organized into relatively distinct fields with their own methods, standards, problem structures, and criteria of legitimate contribution (Whitley, 2000). A technique that is highly valued in one field may be viewed as peripheral, premature, or insufficiently grounded in another. Therefore, the returns to AI should depend on the epistemic and evaluative environment into which AI knowledge is introduced.

The sociology of science has long emphasized that scientific recognition is not determined only by the technical content of research, but also by the norms and audiences of scientific communities (Merton, 1973). Similarly, institutional theory suggests that new practices gain value only when they are perceived as appropriate and legitimate within a given social context (Suchman, 1995). Applying this logic to AI-enabled research, the same act of using AI may carry different meanings across fields. This field-level variation is important because AI knowledge does not enter in a vacuum; instead, it must pass through disciplinary filters. Thus, the value of AI knowledge incorporation is expected to vary across different fields. This leads to the following hypothesis:

**Hypothesis 2.** The citation returns to AI references vary across scientific fields.

#### *2.2.2. Individual-level career stage: signaling and substantive engagement*

The returns to AI references may also vary across scholars' career stages. Researchers at different career stages possess different levels of resources and accumulated reputation and face different professional incentives, which may shape their willingness to pursue novel or risky research directions (Merton, 1968; Packalen and Bhattacharya, 2019; Zuckerman and Merton, 1972). Therefore, AI references may serve different functions for researchers at different stages of their careers.

For senior scholars, they typically possess accumulated scientific capital, broader networks, and greater visibility. Referring AI may function partly as a signal of intellectual renewal for them. In the Mertonian reward system, recognition accumulates unevenly, and established scholars often receive greater visibility for similar contributions because of their prior reputation (Merton, 1968). When senior scholars incorporate AI knowledge, audiences may interpret this behavior as evidence that already recognized researchers are engaging with a frontier area. Because they already possess credibility, senior scholars may receive benefits even from extensive AI adoption, namely the simple presence of AI references.

For junior scholars, who often face stronger pressure to establish distinct expertise and demonstrate competence in emerging areas, however, the symbolic value of merely referring AI may be weaker. Career-stage research has long shown that scientific careers are shaped by age, timing, and accumulated recognition (Blume, 1993; Cui et al., 2026; Zuckerman and Merton, 1972). Without comparable accumulated reputation, young scholars may therefore need to build credibility through more substantive engagement with frontier knowledge rather than superficial association.

This distinction suggests two different mechanisms. For senior scholars, AI references may work as a signal of adaptation and continued relevance, while for junior scholars, AI references may work as a signal of technical capability and frontier engagement. Therefore, extensive AI referencing and intensive AI referencing should have different returns across career stages. Accordingly, we propose the hypotheses:

**Hypothesis 3a.** The citation return to extensive AI referencing is stronger for senior scholars than for junior scholars.

**Hypothesis 3b.** The citation return to intensive AI referencing is stronger for junior scholars than for senior scholars.

This logic also implies that junior scholars should differ in the type of AI knowledge they use. If young researchers should use AI references to demonstrate frontier engagement, they should be more likely to cite recent AI papers, which may indicate closer proximity to the technological frontier and high-quality AI papers, which may provide more credible knowledge inputs (Cui et al., 2026). Thus, we propose:

**Hypothesis 3c.** Junior scholars are more likely to cite newer and higher-quality AI papers.

#### 2.2.3. *Institution-level AI capability: technological divide and translational capacity*

Institutions differ greatly in their AI capability. The Matthew effect suggests that actors with existing advantages are more likely to receive additional recognition and resources (Merton, 1968). Similarly, absorptive capacity theory argues that prior related knowledge improves an organization's ability to recognize, assimilate, and exploit external knowledge (Cohen and Levinthal, 1990). Both perspectives suggest a positive relationship between AI capability and the returns to AI-enabled research.

However, AI-enabled research requires more than just technical AI capability. Research on digital inequality suggests that the benefits of new technologies depend not only on access, but also skills, usage, and the ability to convert technological resources into valued outcomes (DiMaggio and Hargittai, 2001; van Deursen and Helsper, 2015). To create broad scientific impact, AI knowledge must be translated across knowledge boundaries. Prior research on knowledge boundaries emphasizes that knowledge transfer across communities often requires translation, transformation, and boundary-spanning work rather than simple transmission (Carlile, 2003; Tushman, 1977).

This requires a balance between technical understanding and domain relevance. Institutions with weak AI capability may lack the expertise needed to understand and integrate AI knowledge substantively. Their AI references may remain broad but superficial. By contrast, institutions with stronger AI capability may produce AI-enabled papers that are technically sophisticated but also more closely aligned with the internal logic of AI research, thus may not necessarily generate the broadest returns across other scientific fields.

This suggests that the relationship between institutional AI capability and citation returns may be non-linear. Institutions with intermediate AI capability may occupy a particularly advantageous position. They possess enough AI expertise to use AI knowledge meaningfully, but they remain closely connected to the problems, audiences, and evaluation standards. In this sense, their advantage lies not simply absorptive capacity, but translational capacity.

This argument modifies a simple Matthew-effect prediction. Prior advantage matters, but the largest returns may not accrue to those with the greatest AI capability, Instead, the highest returns may emerge where AI expertise and field translation are best balanced. We therefore propose:

**Hypothesis 4.** The relationship between institutional AI capability and the citation returns to AI references is non-linear, with intermediate AI-capability institutions obtaining the largest returns.

### *2.3. Institutional pathways: Before, during, and after publication*

To further unpack the argument proposed in the preceding section, we examine three connected pathways through which institutional AI capability may shape the value of AI-enabled research. These pathways correspond to three stages in the life of a paper: how AI knowledge is used before publication, where the paper is positioned in the knowledge space at publication, and how the paper diffuses after publication.

#### *2.3.1. Before publication: Granularity of AI knowledge use*

Absorptive capacity theory suggests that actors with stronger prior related knowledge are better able to identify and use more complex external knowledge (Cohen and Levinthal, 1990). Similarly, research on technological search argues that organizations search more effectively in knowledge areas where they already possess expertise (Rosenkopf and Nerkar, 2001). Applied to AI-enabled research, institutions with stronger AI capability should be more capable of navigating the detailed structure of AI knowledge. They are more likely to recognize specialized AI topics, evaluate their relevance, and incorporate them into scientific work. Nevertheless, institutions with weaker AI capability may rely more heavily on general AI concepts. It reflects the fact that the cognitive and technical costs of using fine-grained AI knowledge are higher. Therefore, different institutional tiers should also be associated with different granularity of AI knowledge use.

**Hypothesis 5.** Institutions with stronger AI capability draw on more fine-grained AI knowledge.

#### *2.3.2. At publication: Epistemic position*

A second pathway concerns where AI-enabled papers are positioned in the knowledge space at publication. Institutional AI capability may not only affect what kinds of AI knowledge are used, but also shapes the epistemic location of the resulting work. Research on absorptive capacity and technological search shows that organizations often search more effectively, but also locally, in knowledge areas where they already possess expertise (Cohen and Levinthal, 1990; Rosenkopf and Nerkar, 2001). Thus, institutions with stronger AI capability may therefore be better able to integrate specialized AI knowledge, but their papers are also more likely to remain close to AI-centered knowledge spaces.

This position has ambiguous implications for scientific impact. Proximity to AI can reflect deeper technical integration, but it may also reduce distinctiveness as many papers cluster in similar AI-intensive areas. Strong AI capability thus brings both technical centrality and greater competition for attention (Chai and Menon, 2019). We therefore propose:

**Hypothesis 6.** Papers from top AI institutions are more likely to occupy AI-centered epistemic positions, as reflected in greater proximity to cited AI sources, denser semantic neighborhoods, and more AI-intensive local neighborhoods.

#### *2.3.3. After publication: diffusion structure and audience composition*

A third pathway concerns how AI-enabled papers diffuse after publication. Papers from top AI institutions are more likely to be embedded in AI-centered knowledge spaces. Because knowledge flows more easily among cognitively proximate communities, these papers may attract downstream citations disproportionately from AI-related audiences (Boschma, 2005; Nooteboom et al., 2007). They may also affect the structure of AI knowledge diffusion in another way. Because these papers carry stronger technical authority, they may become substitute gateways to the AI sources they cite. Later papers may cite the focal paper without directly citing the original AI references (Funk and Owen-Smith, 2017; Wu et al., 2019). In this case, the focal paper does not merely transmit AI knowledge outward; it also partially replaces the original AI sources in downstream citation chains.

By contrast, intermediate AI-capability institutions may follow a different diffusion pathway. Their AI-enabled papers may be more legible and useful to researchers across non-AI fields. Prior research on knowledge brokering and boundary spanning suggests that actors located between knowledge communities can recombine ideas and make them accessible across otherwise separated audiences (Fleming and Waguespack, 2007; Hargadon and Sutton, 1997). More broadly, knowledge crossing disciplinary or organizational boundaries often requires translation rather than simple transfer (Carlile, 2004). These papers may therefore not be the most technically AI-centered, but they may be better positioned to connect AI knowledge to diverse scientific problems, audiences, and evaluation standards.

We therefore propose:

**Hypothesis 7a.** AI-enabled papers from top AI institutions are more likely to be cited by AI-related audiences.

**Hypothesis 7b.** AI-enabled papers from top AI institutions are more likely to displace the original AI sources in downstream citation chains.

**Hypothesis 7c.** AI-enabled papers from intermediate AI-capability institutions are more likely to be cited by a more diverse range of scientific audiences.

# 3. Methodology

## *3.1. Data*

Our bibliometric data are drawn from SciSciNet V2, a large-scale data lake constructed primarily from the Microsoft Academic Graph. It covers more than 134 million scientific publications and provides standardized information on papers, authors, institutions, and citation relationships (Lin et al., 2023).

Our empirical sample is constructed in three steps. First, we use CSRankings (Berger, 2020) to identify institutions with sustained prior AI capability, retaining those listed in its AI-related fields at least eight years between 2015 and 2025. We then retrieve all publications affiliated with these institutions from SciSciNet V2 for the period 2000-2024.

Second, based on SciSciNet V2's level-0 field classification, which refer to 19 top-level disciplines, we restrict the focal sample to non-computer-science publications, thereby focusing on the use of AI knowledge beyond its field of origin (Zhao and Li, 2026).

Third, we identify AI-related references using OpenAlex's classification labels (Priem et al., 2022). OpenAlex organizes scientific works into a hierarchical taxonomy of fields, subfields, and topics; we assign each paper to a subfield based on the associated primary topic (Priem et al., 2022). A focal paper is treated as AI-referencing if at least one paper in its reference list is classified as Artificial Intelligence subfield in OpenAlex, which has 77 topics (full list provided in **Table S1**). This procedure produces a paper-level measure of AI knowledge use based on observable citation links.

We further clean the data by removing records with missing metadata. The final dataset contains 360 institutions, and approximately 12.8 million focal papers.

## *3.2. Measures*

### *3.2.1. Dependent variable: scientific impact*

The dependent variable of this study is the scientific impact of a focal paper. We measure scientific impact using up-to-five-year citation counts, denoted as $C5$. Prior research has widely used citation counts as a proxy for scientific impact, as citations reflect the extent to which a paper influences subsequent research (Bornmann and Daniel, 2008; Garfield, 1972).To account for differences in citation opportunities across publication years, all models include publication-year fixed effects.

### *3.2.2. Key explanatory variables: AI knowledge integration*

The key explanatory variables capture the extent to which a focal paper integrates AI knowledge through its references. For each focal paper $p$, we define its reference set as $R_p$. Based on the AI-related literature identified above, we define the set of AI-related references cited by $p$ as $A_p = R_p \cap AI$. Where $AI$ denotes the set of papers classified under the OpenAlex subfield of Artificial intelligence. We use this set to construct several measures of AI knowledge integration.

First, we construct an AI reference dummy, $AIRef_p$, which equals one if $p$ cites at least one AI-related paper and zero otherwise. This variable captures the extensive margin of AI knowledge integration.

$$AIRef_p = I(|A_p| > 0). \quad (1)$$

Second, we measure the intensity of AI knowledge integration using the share of AI-related references among all references:

$$AI\ score = |A_p|/|R_p|. \quad (2)$$

Previously examined and validated by Liu et al. (2025), this score allows us to distinguish papers that merely cite AI from that rely more intensively on AI knowledge.

#### *3.2.3. Author career stage*

To examine how the returns to AI knowledge integration vary across scholars at different career stages, we measure each author's academic age, which is defined as the difference between the publication year of the focal paper and the author's first observed publication year in SciSciNet V2.

Because first authorship commonly signals primary responsibility for the research and manuscript in many fields, we use the first author's academic age as the paper-level measure of career stage. This approach reduces the influence of coauthor composition on what is intended to capture an individual-level attribute. Following Wu et al. (2024), we further divide scholars into three groups according to their career stage. The thresholds are defined as follows: early-career (0-5 years), mid-career (6-17 years), and senior (18-50 years). Authors with academic ages exceeding 50 years are excluded to reduce potential errors in author disambiguation and publication-history measurement.

### 3.2.4. *Institutional AI capability*

To examine how institutional capability shapes the returns to AI knowledge integration, we construct a measure of institutional AI capability based on CSRankings (Berger, 2020). For each retained institution, we calculate its average AI ranking across the years in which it appears in CSRankings. Since a lower rank indicates stronger AI capability and ranking data are inherently non-linear—with larger differences among top-ranked institutions and smaller differences among lower-ranked ones—we transform the average rank using the negative logarithm:

$$AICapability_j = -\log\left(\overline{Rank_j}\right). \quad (3)$$

For paper-level analyses, each focal paper is assigned an institutional AI capability measure based on the first author's institution.

To examine potential non-linear effects, we divide papers into four groups using paper-weighted quantile bins. Specifically, focal papers are sorted according to the AI capability of their associated institution and divided into four groups with approximately equal numbers of papers. Q1 represents papers from institutions with the lowest AI capability, while Q4 represents papers from institutions with the highest AI capability. This paper-weighted grouping is appropriate because the unit of analysis is the focal paper and balanced bins improve comparability across groups in paper-level regressions.

### 3.2.5. *Mechanism variables*

**Granularity of AI knowledge use**

We first measure the granularity of AI knowledge used by each focal paper. The theoretical argument suggests that institutions with stronger AI capability are more likely to cite more specialized and fine-grained AI knowledge. To operationalize this idea, we use the hierarchical structure of the Computer Science Ontology (CSO). CSO is a comprehensive, automatically generated ontology comprising approximately 14,000 research concepts and their hierarchical relationships, offering a more detailed structure than general-purpose taxonomies (Salatino et al., 2020). Further, we utilized the pre-computed, high-quality topic alignments provided by the Science Data Lake (Wilinski, 2026), which developed an embedding-based alignment framework using BGE-large sentence embeddings to map OpenAlex topics to 13 scientific ontologies, including CSO.

For each cited AI paper $a$, we calculate its topic depth based on the CSO topics associated with it. For a focal paper $p$, the granularity of AI knowledge use is measured as the average depth of the AI papers it cites:

$$AIGranularity_p = \sum_{a \in A_p} Depth_a / |A_p|, \tag{4}$$

where $Depth_a$ denotes the CSO-based topic depth of $a$; when a topic appears in multiple positions within the CSO, we use the maximum depth among those positions. Higher values indicate that the focal paper draws on more-fine-grained and specialized AI knowledge.

**Epistemic crowding**

To capture the semantic content of scientific papers, we utilized the pre-computed document embeddings provided by the SciSciNet V2 dataset (Lin et al., 2023). These embeddings were generated using the SPECTER2 model (Singh et al., 2023), which is designed to represent scientific documents as vectors in a 768-dimensional space based on abstracts and citation patterns.

We construct two measures of epistemic crowding. Facebook AI Similarity Search (FAISS), a Python library, is used for efficient vector similarity search (Douze et al., 2025). For each focal paper, we identify its $k$ nearest neighbors $NN_k(p)$ in the embedding space. In the main analysis, we set $k = 50$.

Our first measure is the average cosine distance between focal paper $p$ and its $k$ nearest neighbors. Lower values indicate that $p$ is located in a denser semantic neighborhood:

$$NeighborDistance_p = \frac{1}{k} \sum_{q \in NN_k(p)} \left(1 - \frac{\mathbf{e}_p^{\top} \mathbf{e}_q}{\|\mathbf{e}_p\| \|\mathbf{e}_q\|}\right), \tag{5}$$

Where $\mathbf{e}_p$ and $\mathbf{e}_q$ denotes embeddings of $p$ and $q$, respectively.

The second measure captures the AI intensity of the local neighbors:

$$NeighborAIShare_p = \frac{1}{k} \sum_{q \in NN_k(p)} AIRef_q. \tag{6}$$

**AI-source displacement**

Finally, we construct an AI-source displacement measure to examine whether a focal paper becomes a new gateway to the AI knowledge it cites. The logic follows disruption-based measures in citation networks (Funk and Owen-Smith, 2017; Wu et al., 2019).

For each focal paper $p$, we define its cited AI-source set $A_p$. We then examine downstream citation behavior within five years after the publication of the focal paper. Subsequent papers can be divided into three categories: papers that cite the focal paper but not its original AI sources, papers that cite both the focal paper and its original AI sources, and papers that cite the original AI sources but not the focal paper. We denote the counts of these three categories as $N_F, N_B$ and $N_R$ respectively. The AI-source displacement index is then calculated as:

$$AISourceDisplacement_p = \frac{N_F - N_B}{N_F + N_B + N_R}. \quad (7)$$

A higher value indicates that downstream papers are more likely to cite the focal paper while bypassing the original AI sources it cited, which suggests that the focal paper has become a substitute gateway through which AI knowledge is accessed and reused. We also calculate the average cosine distance of focal paper to its AI papers cited.

**Audience composition**

We further characterize the audience composition of each focal paper by examining the papers that cite it within five years after publication. While citation counts measure the overall level of scientific impact, they do not reveal which scientific communities pay attention to the focal paper. To capture this dimension, we construct two audience-based measures: the breadth of downstream audiences and the size of the AI-related audience.

First, we measure the breadth of downstream audiences using the subfield distribution of citing papers. For each focal paper $p$, let $C_p$ denotes the set of papers that cite $p$ within the five -year citation window. Each citing paper is assigned to an OpenAlex subfield. We calculate the share of citing papers from each subfield $s$ as $\pi_{ps}$. Then we use Shannon entropy to measure the diversity of the focal paper's audience:

$$AudienceEntropy_p = -\sum_s \pi_{ps} \log(\pi_{ps}). \quad (8)$$

Higher values indicate that the focal paper is cited by a more diverse set of scientific subfields, suggesting broader cross-field diffusion.

We also count the size of the AI-related audiences, defined as the number of papers in $C_p$ that are classified under the OpenAlex subfield of Artificial intelligence.

### 3.3. *Regression model*

Citation counts are non-negative, highly skewed, and include a large number of zero-citation observations. Thus, we use Poisson pseudo-maximum likelihood (PPML) models as the main specification (Santos Silva and Tenreyro, 2011; Silva and Tenreyro, 2006). Our baseline model is specified as follows:

$$E(C5_p|\cdot) = \exp\left(\alpha + \beta_1 AIRef_p + \beta_2 AI\ score_p + \mathbf{X}_p'\boldsymbol{\gamma} + \boldsymbol{FEs}\right), \quad (9)$$

where $E(C5_p|\cdot)$ denotes the conditional mean of $C5_p$ given the explanatory variables, controls, and fixed effects included on the right-hand side. $\mathbf{X}_p$ is a vector of control variables, including count of references (in logarithm), team size, first-author's h-index and institution count. $\boldsymbol{FEs}$ denotes the fixed effects included in the model, including publication-year fixed effects, field fixed effects and document-type fixed effects. Standard errors are clustered at the institution level to account for within institution correlation.

### 3.4. *Exploratory coefficient-attenuation analysis*

To examine which candidate pathways are most closely associated with differences in the returns to intensive AI referencing across institutional capability groups, we conduct an exploratory coefficient-attenuation analysis. For each mechanism variable $M$, we estimate a baseline model and an augmented model on the AI-referencing papers. The baseline PPML model is specified as follows:

$$E(\,C5_p\mid\cdot\,) = \exp\left(\alpha + \beta_0 AI\ score_p + \sum_{i=2}^{4}(\beta_i Qi_p + \delta_i Qi_p \times AI\ score_p) + \mathbf{X}_p'\boldsymbol{\gamma} + \boldsymbol{FEs}\right), \tag{10}$$

where $Qi_p$ is an indicator variable equal to one if paper $p$ is assigned to AI-capability group $Qi$.

The augmented model additionally includes one candidate mechanism variable at a time. We then calculate the differences of the $\delta_i$'s before and after adding the mechanism variables. One-sided Wald tests are used to evaluate the significances of the group differences in AI-score returns. Because some mechanism variables are measured after publication, this analysis is not interpreted as a causal mediation test. Instead, it provides descriptive evidence on which characteristics are most closely associated with the intermediate-capability advantage.

### *3.5. Statistical tests*

In addition to the regression analyses, we use statistical tests to examine group differences in several mechanism variables. Because the AI-capability groups may differ in both sample size and variance, we use Welch's two-sample *t*-tests to compare group means (Welch, 1947). For variables involving comparisons across the four AI-capability groups, we conduct pairwise tests and apply Bonferroni correction to account for multiple comparisons, with Bonferroni-adjusted $p$-values reported in the results (Dunn, 1961).

# 4. Results

### *4.1. The prevalence and uneven diffusion of AI in science*

We first examine the extent to which AI-related knowledge has entered science. **Figure 2** presents descriptive evidence on the prevalence and intensity of AI references across SciSciNet level-0 fields. The left panel reports the share of papers with a non-zero AI score. The right panel shows the distribution of AI scores among papers with non-zero AI scores, with the top 1% of values trimmed within each field.

The left panel shows that AI have become visible across a wide range of scientific fields. However, as the right panel shows, the prevalence of AI knowledge integration varies by field. Mathematics has the highest share, followed by Geology and Psychology. By contrast, Medicine, Material science and Chemistry show much lower levels of AI-referencing papers. The right panel shows that the intensity of AI knowledge is highly skewed. The median AI score (marked by the red line) is relatively low, while the distribution has a long right tail.

**Figure 2 Diffusion of AI-related references across scientific fields.**

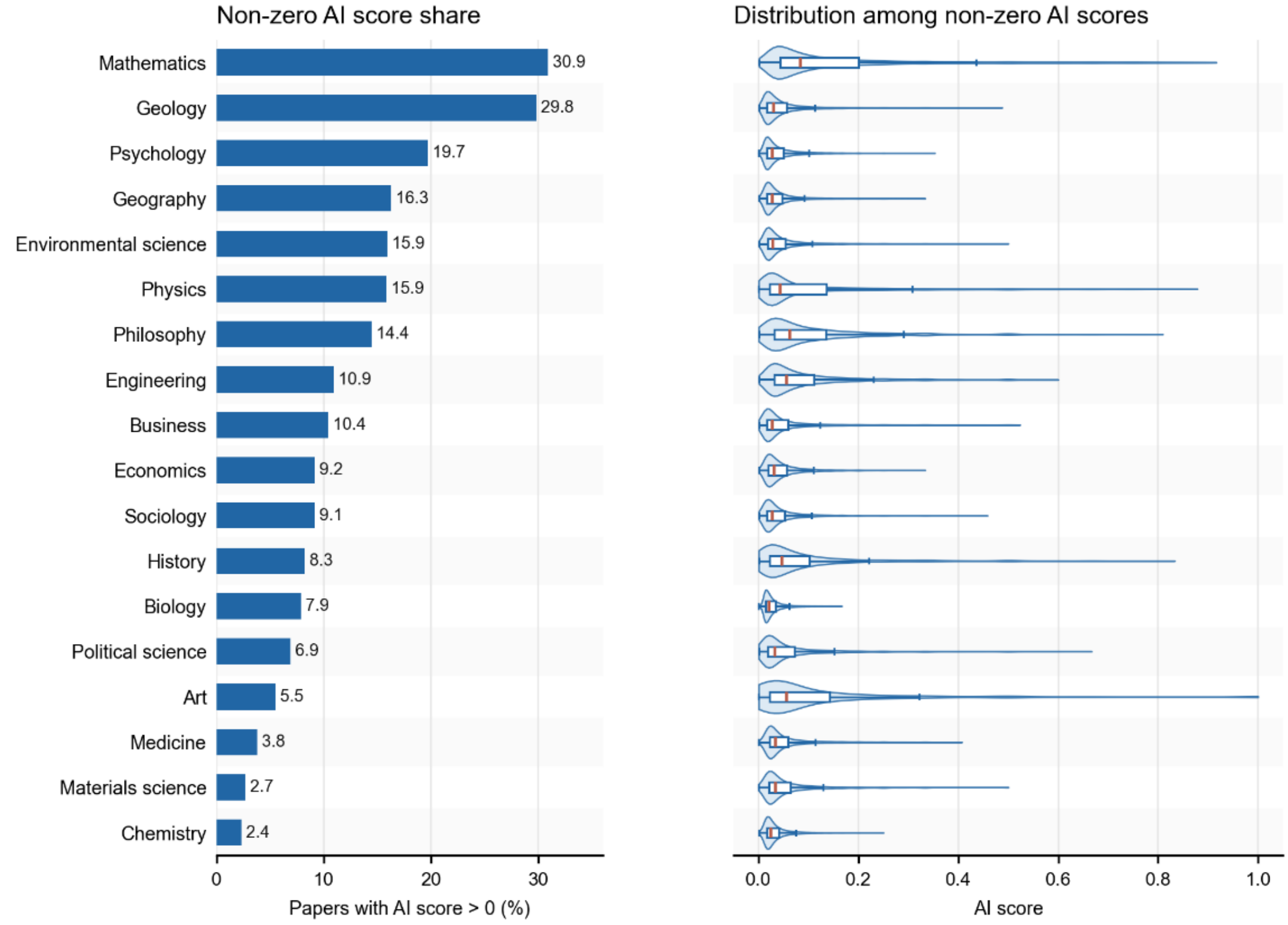

*Notes.* The figure summarizes cross-field variation in the use of AI knowledge. The left panel shows the percentage of focal paper with a non-zero AI score, while the right panel shows the distribution of AI score among focal papers with non-zero AI scores. The violin plots represent the estimated distributions, with the top 1% trimmed to improve readability. The embedded box plots indicate the medians and interquartile ranges.

## 4.2. *AI boosts scientific impact*

错误!未找到引用源。 reports the regression results for the relationship between AI knowledge integration and five-year citation impact. Descriptive statistics for the main variables and their pairwise Pearson correlations are reported in **Table S2** and **Table S3,** respectively. As a robustness check, **Table S4** reports OLS estimates using $\ln(C5+1)$ as the dependent variable.

**Hypothesis 1a** predicts that paper citing AI-related references receive higher citation impact. The result provides general support for this expectation. In Column (1), the coefficient of AI dummy is positive and statistically significant ($\beta = 0.063, p < 0.01$), indicating that papers with at least one AI-related reference receive more five-year citations. The coefficient remains positive and significant in Column (3) after AI score is included.

**Hypothesis 1b** predicts that intensive AI knowledge integration is positively associated with citation impact. The results strongly support this hypothesis. In Column (2), the coefficient of AI score is positive and statistically significant ($\beta = 0.667, p < 0.01$). In Column (3) The result remains. These results indicate that papers incorporating more AI knowledge tend to receive higher citation impact.

Column (5) further examines the potential nonlinear effect of AI knowledge integration. The coefficient of AI score is positive and significant, whereas the coefficient of its squared term is negative and significant. This result suggests diminishing marginal returns to AI knowledge integration. In other words, incorporating AI knowledge is generally associated with higher citation impact, but the citation advantage does not increase indefinitely as the share of AI-related references becomes very high. Overall, these findings well support **Hypothesis 1a**, **Hypothesis 1b** and **Hypothesis 1c**.

**Table 1 The Impact of AI References on Citations**

| | (1) | (2) | (3) | (4) | (5) |
|---|---|---|---|---|---|
| | C5 | C5 | C5 | C5 | C5 |
| AIRef | 0.063*** | | 0.032*** | | 0.007 |
| | (0.008) | | (0.008) | | (0.007) |
| AI score | | 0.667*** | 0.571*** | 1.392*** | 1.321*** |
| | | (0.043) | (0.036) | (0.130) | (0.118) |
| AI score$^2$ | | | | -1.268*** | -1.180*** |
| | | | | (0.178) | (0.162) |
| ln(refs count+1) | 0.602*** | 0.606*** | 0.604*** | 0.606*** | 0.606*** |
| | (0.009) | (0.009) | (0.009) | (0.009) | (0.009) |
| Team size | 0.075*** | 0.075*** | 0.075*** | 0.075*** | 0.075*** |
| | (0.001) | (0.001) | (0.001) | (0.001) | (0.001) |
| h-index | 0.009*** | 0.009*** | 0.009*** | 0.009*** | 0.009*** |
| | (0.000) | (0.000) | (0.000) | (0.000) | (0.000) |
| Institution count | 0.014*** | 0.014*** | 0.014*** | 0.014*** | 0.014*** |
| | (0.002) | (0.002) | (0.002) | (0.002) | (0.002) |
| Constant | 0.228*** | 0.215*** | 0.219*** | 0.214*** | 0.215*** |
| | (0.042) | (0.041) | (0.042) | (0.041) | (0.042) |
| Observations | 12827989 | 12827989 | 12827989 | 12827989 | 12827989 |
| Pseudo $R^2$ | 0.359 | 0.359 | 0.359 | 0.359 | 0.359 |

*Notes:* 1. Clustered standard errors at the institution level are reported in parentheses; 2. * $p < 0.1$, ** $p < 0.05$, *** $p < 0.01$

### *4.3. Field heterogeneity in the returns to AI references*

**Hypothesis 2** predicts that the citation returns to AI references vary across fields. **Figure 3** provides strong support for this hypothesis by estimating field-specific associations. The left panel reports the estimated percent effects of the $AIRef$, while the right panel reports the estimated percent effects of 0.01 increase in $AI\ score$.

For the extensive margin, the $AIRef$ is positively associated with citation impact in several fields, such as Engineering, Philosophy, Sociology. These results suggest that, in these fields, simply citing at an AI-related paper is associated with higher impact. However, the effect is negative in several other fields. This suggests that the presence of AI references alone does not always generate citation advantages.

The right panel shows that the intensive margin also varies. In many fields, $AI\ score$ is positively associated with citation impact, indicating that papers with a higher share of AI-related references tend to receive greater scientific impact. Even in fields where mere citing AI is not always rewarded, more intensive incorporation of AI knowledge may still be associated with higher citation impact. At the same time, $AI\ score$ is negatively associated with citation impact in some fields, including Engineering, Economics, and Materials science. This contrast suggests that intensive AI referencing may not always be beneficial.

Taken together, these patterns are consistent with the idea that the value of AI knowledge depends on field-specific audiences, evaluation standards, and expectations about legitimate research practices. Thus, **Hypothesis 2** is supported.

**Figure 3 Extensive margin and intensive margin across fields.**

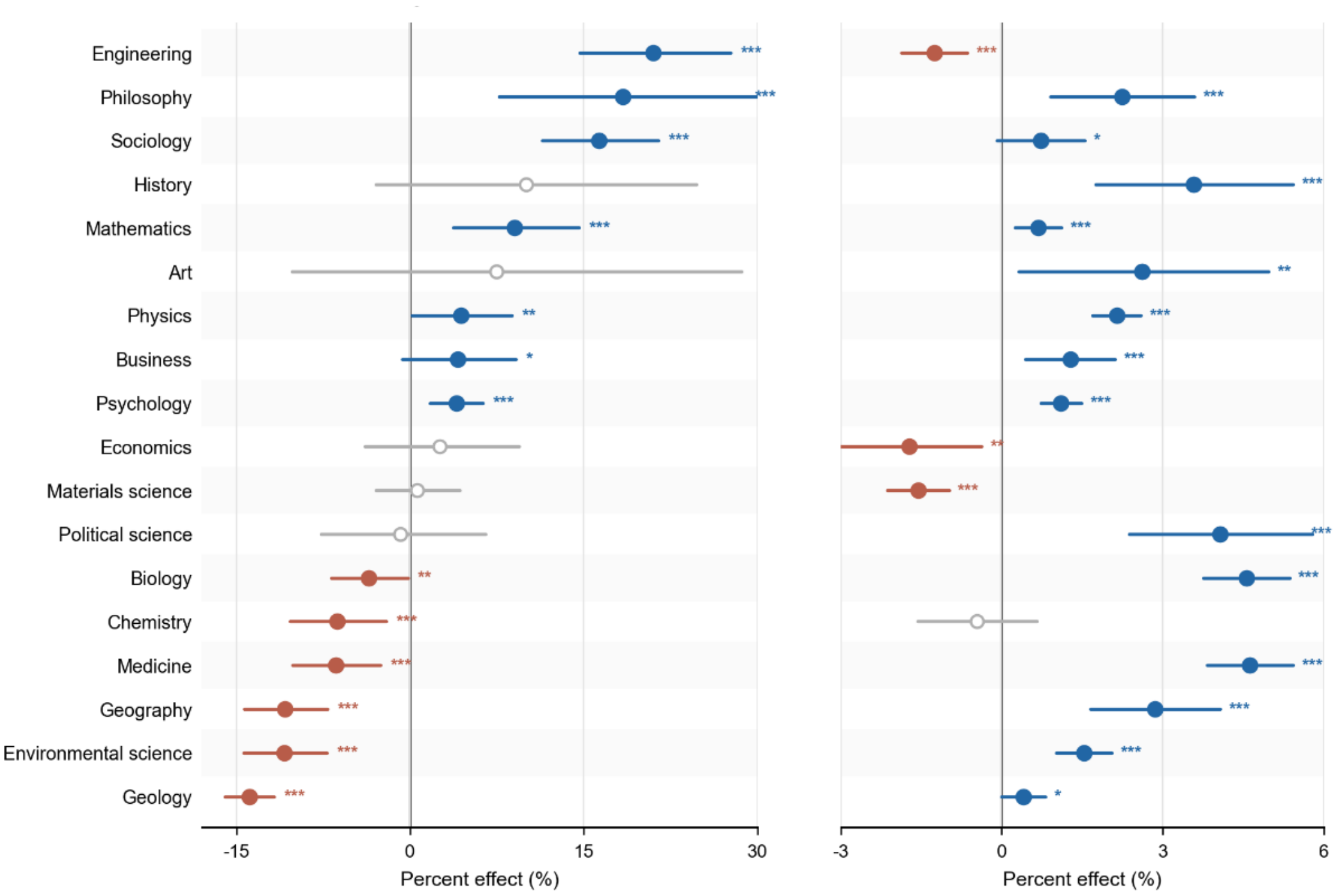

*Notes.* This figure reports field-specific estimates of the association between AI knowledge integration and five-year citation impact. The left panel presents the estimated percentage change in five-year citations associated with $AIRef$, calculated as $100[\exp(\beta) - 1]$. The right panel presents the estimated percentage change associated with a 0.01 increase in the $AI\ score$, calculated as $100[\exp(0.01\beta) - 1]$. Circles indicate point estimates, and horizontal lines represent 95% confidence intervals. Blue and red markers denote statistically significant positive and negative estimates, respectively, while gray markers denote estimates that are not statistically significant. The vertical line indicates a null effect. *, **, and *** indicate statistical significance at the 10%, 5%, and 1% levels, respectively.

### *4.4. Career stage and the returns to AI references*

Hypothesis 3a and 3b examine whether the citation returns differ across scholars at different career stages. Table 2 reports PPML estimates by first-author academic age group. Column (1) presents the interaction model using early-career scholars as the reference group. The coefficient of the $AIRef$ is negative and statistically significant ($\beta = -0.0215,\ p < 0.01$), indicating that, among early-career scholars, simply citing AI is not rewarded. However, the interaction terms between the $AIRef$ and older career-stage groups are positive and statistically significant. The coefficient for $AIRef \times 5 < Age \leq 17$ is positive ($\beta = 0.0265, p < 0.01$), and the coefficient for $AIRef \times Age > 17$ is even larger ($\beta = 0.0352, p < 0.01$). These results suggest that the citation return to extensive AI referencing increase with academic age. The results for AI score show the opposite pattern. In Column (1), the coefficient of AI score is positive and statistically significant ($\beta = 0.5345, p < 0.01$), indicating that more intensive AI knowledge integration is associated with higher citation impact among early-career scholars. The coefficient for $AI\ score \times 5 < Age \leq 17$ is negative and weakly significant ($\beta = -0.0554, p < 0.10$), while the coefficient for $AI\ score \times Age > 17$ is also negative and statistically significant ($\beta = -0.0806, p < 0.05$). These results indicate that the citation return to intensive AI referencing declines with academic age. The subgroup regressions in Column (2)-(4) show the same pattern.

Taken together, these results suggest that career stage shapes how AI knowledge integration is rewarded. Possibly, for senior scholars, AI references work as a signal, and for junior scholars, they must work as substance. Overall, **Hypothesis 3a** and **Hypothesis 3b** are both supported.

**Table 2 PPML Regression by Citation Age Group**

| | (1)<br>Interaction | (2)<br>Age <= 5 | (3)<br>5 < Age <= 17 | (4)<br>Age > 17 |
|---|---|---|---|---|
| AIRef | -0.022*** | -0.014** | 0.001 | 0.009* |
| | (0.006) | (0.006) | (0.005) | (0.006) |
| AIRef × 5 < age ≤ 17 | 0.027*** | | | |
| | (0.005) | | | |
| AIRef × age > 17 | 0.035*** | | | |
| | (0.006) | | | |
| AI score | 0.535*** | 0.544*** | 0.476*** | 0.447*** |
| | (0.039) | (0.041) | (0.034) | (0.035) |
| AI score × 5 < age ≤ 17 | -0.055* | | | |
| | (0.033) | | | |
| AI score × age > 17 | -0.081** | | | |
| | (0.037) | | | |
| Age | -0.003*** | 0.027*** | -0.017*** | 0.000 |
| | (0.000) | (0.004) | (0.003) | (0.001) |
| Age$^2$ | -0.000 | -0.005*** | 0.000** | -0.000*** |
| | (0.000) | (0.001) | (0.000) | (0.000) |
| Controls | Yes | Yes | Yes | Yes |
| Year FE | Yes | Yes | Yes | Yes |
| Field FE | Yes | Yes | Yes | Yes |
| Doctype FE | Yes | Yes | Yes | Yes |
| Observations | 11927328 | 3,725,231 | 4,142,149 | 4,059,948 |
| Pseudo $R^2$ | 0.3651 | 0.3558 | 0.3592 | 0.3821 |

Notes: 1. Clustered standard errors at the institution level are reported in parentheses; 2. Controls: ln(reference count+1), Team size, h-index, Institution count and dummy variables of age groups; 3. * $p < 0.1$, ** $p < 0.05$, *** $p < 0.01$.

Figure 4 provides descriptive evidence on how the quantity, recency, and quality of cited AI knowledge vary across career stages. The results support **Hypothesis 3c**. Panel a shows that early-career scholars cite more AI-related papers than senior scholars. Panel b and c indicate that early-career scholars rely more on recent AI knowledge. Panel b reports the average age of cited AI papers; however, because the OpenAlex taxonomy assigns some very old papers to the Artificial intelligence subfield, this mean can be influenced by extreme values. To address this concern, Panel c provides a complementary measure based on the share of AI references published within the past five years. Both panels consistently show that scholars cite more recent AI knowledge. This result echoes the findings of Cui et al. (2026) Panel d shows that junior scholars also cite higher-quality AI papers, as reflected in higher average citation counts. These may help explain their higher returns to AI intensity, but we do not find enough evidence.

**Figure 4 Quantity, recency and quality of cited AI knowledge across career stages.**

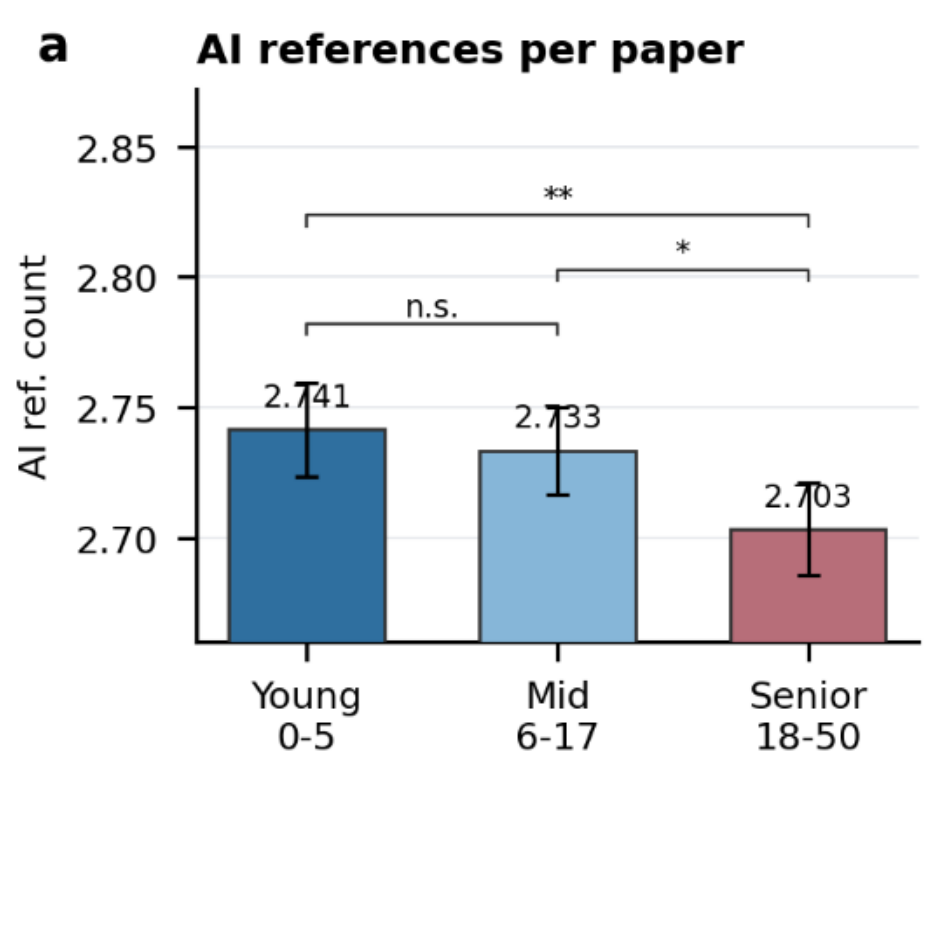

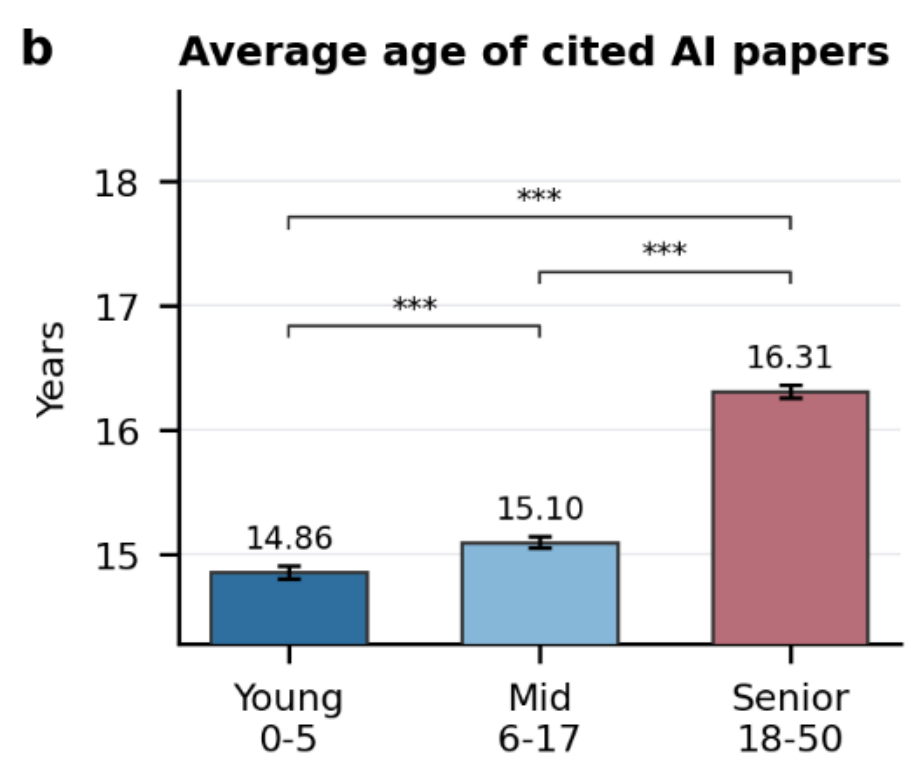

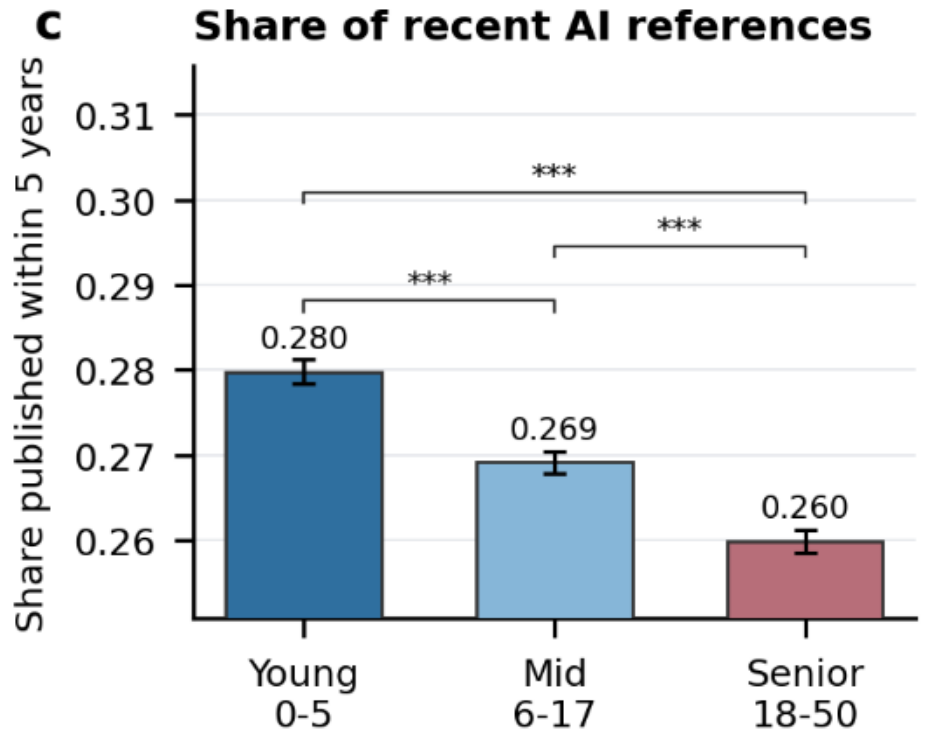

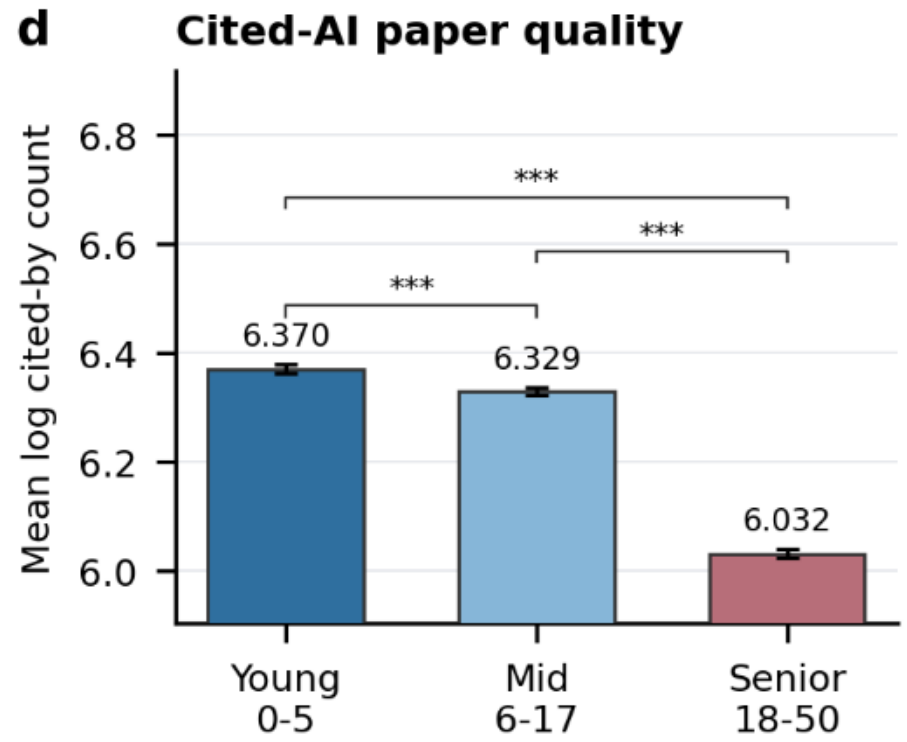

*Notes.* The figure compares patterns of AI knowledge use across career stages. Panel (a) reports the mean number of AI-related references per focal paper. Panel (b) reports the mean age of cited AI papers, measured as the difference between the publication year of the focal paper and that of each cited AI paper. Panel (c) reports the mean share of AI references published within the five years preceding the focal paper. Panel (d) reports the mean quality of cited AI papers, measured by their log-transformed citation counts. Bars indicate group means, and error bars represent 95% confidence intervals. Brackets report pairwise Welch's $t$-tests with Bonferroni-adjusted $p$-values. n.s. denotes a statistically insignificant difference; *, **, and *** indicate significance at the 10%, 5%, and 1% levels, respectively.

### 4.5. *Institutional AI capability and the returns to AI references*

**Hypothesis 4** predicts a non-linear relationship between institutional AI capability and the returns. **Table 3** reports fixed-effects PPML regressions estimated separately for four institutional AI capability groups.

The results show that AI score is positively and significantly associated with five-year citation impact across all four groups. However, the magnitude of the coefficient varies substantially across institutional AI-capability groups. The coefficient is 0.463 for Q1 institutions, increases to 0.663 for Q2 institutions, remains high at 0.525 for Q3 institutions, and declines to 0.501 for Q4 institutions. This pattern indicates that the proportional citation return to AI knowledge integration is largest among intermediate AI-capability institutions rather than among the lowest- or highest-capability institutions.

These findings support the non-linear pattern proposed by **Hypothesis 4**. Institutions with the strongest AI capability do not obtain the largest return. Instead, intermediate institutions appear to benefit more. This suggests that AI capability is important, but more capability does not automatically translate into greater impact.

**Table 3 PPML Regression by institutional AI capability group**

| | (1)<br>Q1 lowest | (2)<br>Q2 | (3)<br>Q3 | (4)<br>Q4 highest |
|---|---|---|---|---|
| AIRef | 0.003 | 0.022* | 0.040*** | 0.047** |
| | (0.008) | (0.013) | (0.014) | (0.020) |
| AI score | 0.463*** | 0.663*** | 0.525*** | 0.501*** |
| | (0.063) | (0.073) | (0.055) | (0.073) |
| Age | -0.011*** | -0.010*** | -0.010*** | -0.014*** |
| | (0.001) | (0.001) | (0.001) | (0.001) |
| $Age^2$ | 0.000*** | 0.000*** | 0.000*** | 0.000*** |
| | (0.000) | (0.000) | (0.000) | (0.000) |
| ln(refs count+1) | 0.586*** | 0.612*** | 0.573*** | 0.631*** |
| | (0.009) | (0.014) | (0.016) | (0.025) |
| Team size | 0.066*** | 0.070*** | 0.076*** | 0.079*** |
| | (0.002) | (0.002) | (0.002) | (0.003) |
| h-index | 0.014*** | 0.013*** | 0.013*** | 0.013*** |
| | (0.000) | (0.000) | (0.000) | (0.000) |
| Institution count | 0.022*** | 0.011*** | 0.011*** | 0.001 |
| | (0.00623) | (0.00593) | (0.00651) | (0.00781) |
| Observations | 3,061,628 | 2,923,136 | 3,052,956 | 2,904,616 |
| pseudo $R^2$ | 0.3701 | 0.3701 | 0.3567 | 0.3499 |

*Notes.* 1. Dependent variable: C5; 2. Clustered standard errors at the institution level are reported in parentheses; 3. $^{*}$ $p < 0.10$, $^{**}$ $p < 0.05$, $^{***}$ $p < 0.01$.

As a robustness check, **Table S5** reclassifies institutional AI capability into three groups, yielding results consistent with the main analysis.

### *4.6. Profiling knowledge use, semantic position, and scientific audiences*

The preceding results show that the proportional citation return is largest among intermediate AI-capability institutions, to better understand this pattern, we further examine how papers from different institutional AI-capability groups differ in their knowledge use, semantic position, and downstream reception.

**Table 4** reports mean values of six variables across four groups. Several patterns stand out. First, papers from the highest AI-capability institutions cite more fine-grained AI knowledge. This supports the idea of **Hypothesis 5** that leading AI institutions draw on more specialized AI knowledge.

Second, papers from higher-capability institutions are located in denser and more AI-intensive semantic neighborhoods. The mean distance to the 50 nearest neighbors decreases monotonically from 0.06779 in Q1 to 0.06553 in Q4, indicating that papers from top AI institutions are located in more crowded regions of the knowledge space. At the same time, the share of paper using AI among the 50 nearest neighbors increases from 0.336 in Q1 to 0.374 in Q4. These differences are statistically significance across all adjacent groups. Together, these results suggest that stronger institutional AI capability is associated not only with greater technical depth, but also with stronger embeddedness in AI-intensive epistemic neighborhoods, and possibly, more competitive space. These provide evidence that supports **Hypothesis 6**.

Third, papers from top AI institutions are semantically closer to the AI papers they cite.

The average decrease from 0.12505 in Q1 to 0.12243 in Q4. Since lower values indicate greater semantic proximity, this pattern suggests that Q4 papers are more closely aligned with their cited AI sources.

Fourth, the AI-source displacement measure also increases with institutional AI capability. This suggests that papers from institutions with greater AI capability are more likely to become substitute citation gateways to the AI sources they cite. However, the differences are not all statistically distinguishable. This provides tentative support for the **Hypothesis 7b**.

**Table 4 Profiles across institutional AI-capability groups**

| | Q1 | Q2 | Q3 | Q4 |
|---|---|---|---|---|
| AIGranularity | 4.062[a] | 4.065[a] | 4.073[a] | 4.233[b] |
| | (0.003) | (0.004) | (0.003) | (0.004) |
| NeighborDistance | 0.06779[d] | 0.06684[c] | 0.06641[b] | 0.06553[a] |
| | (0.00002) | (0.00003) | (0.00002) | (0.00003) |
| NeighborAIShare | 0.336[a] | 0.348[b] | 0.356[c] | 0.374[d] |
| | (0.001) | (0.001) | (0.001) | (0.001) |
| AISourceDisplacement | -0.021[a] | 0.004[b] | 0.008[b] | 0.011[b] |
| | (0.002) | (0.002) | (0.002) | (0.002) |
| AudienceEntropy | 0.922[a] | 0.927[ab] | 0.949[c] | 0.927[b] |
| | (0.001) | (0.001) | (0.001) | (0.001) |
| DistanceToCitedAI | 0.12505[c] | 0.12428[b] | 0.12411[b] | 0.12243[a] |
| | (0.00009) | (0.00010) | (0.00009) | (0.00010) |

*Notes.* 1. Cells report the mean values, with standard errors in parentheses; 2. Within each row, means with different superscript letters are significantly different at the 5% level based on Welch two-sample

*t*-tests with Bonferroni correction. Shared letters indicate that the corresponding pairwise difference is not statistically significant after correction.

The audience entropy results reveal a different pattern. Audience subfield entropy is highest for Q3 institutions. This indicates that, in the observed data, papers from Q3 institutions receive citations from a more diverse set of subfields. The higher audience entropy of Q3 papers is consistent with the idea that intermediate-capability institutions may be especially effective at translating AI knowledge into contributions that travel across broader scientific audiences. Therefore, **Hypothesis 7c** is also supported.

For each focal paper, we calculate the share of its five-year citing papers that are classified under the OpenAlex Artificial intelligence subfield. **Figure 5** reports the percentage of focal papers whose AI-subfield citation share reaches a series of thresholds. The pattern is clear: papers from more capable institutions are consistently more likely to receive citations from AI-related audiences, supporting **Hypothesis 7a**. Therefore, the lower broad citation return of Q4 papers should not be interpreted as weaker AI use. Rather, it suggests a different impact pathway: top AI institutions are more likely to produce AI-enabled research that circulates back into AI-related communities, whereas intermediate-capability institutions appear more effective at generating broader cross-field citation returns.

**Figure 5 AI-related audience concentration**

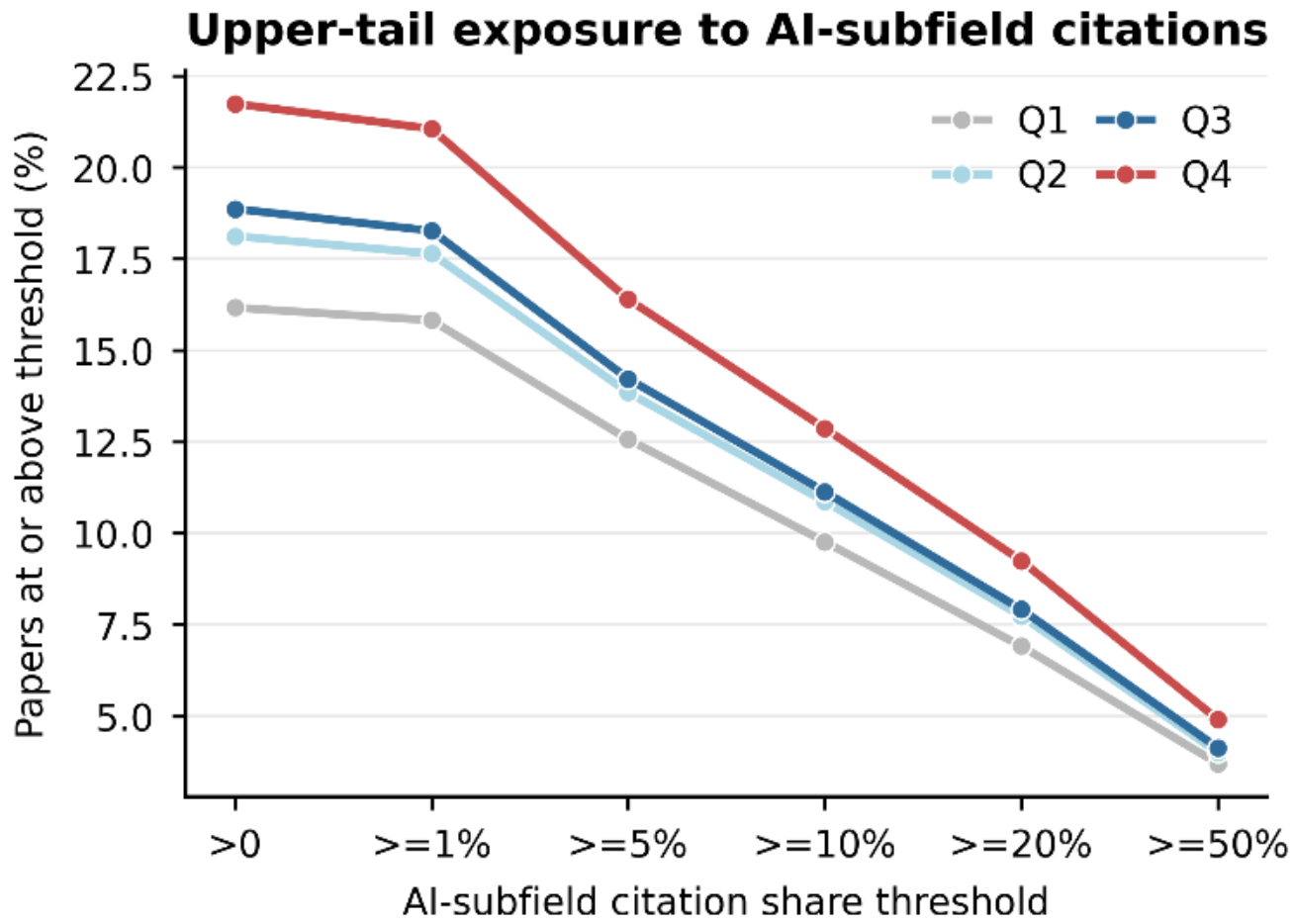

*Notes.* reports the percentage of focal papers whose AI-subfield citation share reaches a series of thresholds, separately by institutional AI capability quartile.

### 4.7. *Coefficient-attenuation results*

**Table 5** shows the estimating results. $AudienceEntropy$ produces the most attenuation. Adding it reduces the Q2-Q3 gap from 0.222 to 0.170, corresponding to a 23.5% reduction, and reduces the Q3-Q4 gap from 0.192 to 0.134, corresponding to a 30.3% reduction. The Wald tests also become insignificant ($p$>0.1) after $AudienceEntropy$ is added. This suggests that broader downstream audience composition is closely associated with the larger AI-score returns of intermediate-capability institutions. In other words, papers from Q2 and Q3 institutions appear to benefit partly because their AI-related knowledge integration reaches a more diverse set of scientific subfields.

$NeighborDistance$ also provides evidence for attenuation, but mainly for the Q3-Q4 comparison. This variable reduces the gap by 37.5%. This suggests that the difference between Q3 and Q4 is partly related to the semantic position of their papers. Papers from Q4 institutions are located in denser knowledge neighborhoods, where the marginal citation return to additional AI knowledge may be lower and more competitive.

**Table 5 Exploratory attenuation tests of AI-score return gaps**

| Variable | Comparison | N | Before | After | Reduction | Attenuation |
|---|---|---|---|---|---|---|
| AIGranularity | $\delta_2 - \delta_4$ | 965,784 | 0.216 (0.070) | 0.205 (0.081) | 0.011 | 5.2%= $\frac{\text{Before}-\text{After}}{\text{Before}}$ |
| | $\delta_3 - \delta_4$ | | 0.190 (0.078) | 0.179 (0.089) | 0.011 | 5.8% |
| NeighborDistance | $\delta_2 - \delta_4$ | 778,474 | 0.143 (0.165) | 0.139 (0.157) | 0.005 | 3.2% |
| | $\delta_3 - \delta_4$ | | 0.189 (0.067) | 0.118 (0.165) | 0.071 | 37.5% |
| NeighborAIShare | $\delta_2 - \delta_4$ | 778,474 | 0.143 (0.165) | 0.134 (0.186) | 0.009 | 6.5% |
| | $\delta_3 - \delta_4$ | | 0.189 (0.067) | 0.186 (0.071) | 0.003 | 1.6% |
| AI-source displacement | $\delta_2 - \delta_4$ | 982,515 | 0.214 (0.070) | 0.213 (0.065) | 0.001 | 0.7% |
| | $\delta_3 - \delta_4$ | | 0.190 (0.075) | 0.174 (0.094) | 0.016 | 8.3% |
| AudienceEntropy | $\delta_2 - \delta_4$ | 944,248 | 0.222 (0.056) | 0.170 (0.103) | 0.052 | 23.5% |
| | $\delta_3 - \delta_4$ | | 0.192 (0.056) | 0.134 (0.120) | 0.058 | 30.3% |

*Notes.* 1. Each variable is estimated on its own maximum available sample. 2. *p*-values in parentheses are based on one-sided Wald tests

Other variables show weaker attenuations and do not change the Wald tests' significances. Taken together, these results point to two main patterns. First, audience breadth is the most consistent pathway associated with the higher AI-score returns of intermediate-capability institutions. Second, epistemic crowding helps explain why the highest-capability institutions do not obtain the largest broad citation return. These findings reinforce the central argument of the paper: the scientific value of AI knowledge depends not only on technical depth or AI-centeredness, but also on whether AI knowledge can be translated into contributions that circulate across broader scientific audiences.

## 5. Discussion

### *5.1. Theoretical implications*

This study advances research on AI and the organization of science by shifting attention from the diffusion of AI knowledge to its conversion into scientific recognition. Research on digital inequality has long distinguished access to technology from differences in use and outcomes (DiMaggio and Hargittai, 2001; van Deursen and Helsper, 2015). Our framework extends this distinction to knowledge-intensive organizations: the increasing availability of AI knowledge does not eliminate inequality, but relocates it to the processes through which knowledge is interpreted, legitimized, and rewarded.

The study also highlights the multilevel nature of technological value creation. The value of a general-purpose knowledge input is not determined by its technical properties alone. It is shaped by disciplinary evaluation systems, researchers' positions within status hierarchies, and the knowledge structures of their organizations. This perspective connects research on technological change with the organization of science by showing that emerging technologies enter preexisting systems of legitimacy, reputation, and organizational differentiation.

Finally, the study suggests a distinction between absorptive capacity and translational capacity (Cohen and Levinthal, 1990). Absorptive capacity concerns an organization’s ability to recognize, understand, and use specialized external knowledge. Translational capacity concerns its ability to make that knowledge meaningful and useful beyond the community in which it originated. This distinction qualifies linear accounts of cumulative advantage (Merton, 1968) and suggests that organizational capability may shape the form, rather than simply the quantity, of scientific influence.

### 5.2. Policy implications

The findings also have several implications for science and innovation policy.

First, expanding access to computing resources, technical expertise, and AI infrastructure remains important, but these investments alone may not ensure that AI knowledge generates value across scientific domains. Funding agencies and universities should therefore complement technical capacity building with support for interdisciplinary collaboration, shared methodological services, and organizational roles that connect AI specialists with domain researchers.

Policy should also recognize that research organizations may contribute to AI-enabled science in different but complementary ways. Institutions at the technological frontier are essential for developing advanced methods, infrastructure, and specialized talent, whereas other institutions may play an important role in adapting AI knowledge to disciplinary problems and extending its reach across scientific communities. Rather than evaluating all institutions against a single model of AI excellence, funding systems may benefit from supporting a portfolio of capabilities and organizational roles.

Finally, the promotion of AI in science should be sensitive to disciplinary and career contexts. Uniform incentives to adopt AI may encourage symbolic association rather than substantive integration. Training and funding programs should therefore be aligned with field-specific research needs and provide junior researchers with sustained opportunities to develop both technical and domain expertise. Evaluation systems should likewise recognize multiple forms of contribution, including technical advancement, domain-specific application, and cross-field knowledge diffusion.

### *5.3. Limitations*

This study has several limitations. First, we use AI references as observable traces of AI knowledge integration. References provide a scalable way to identify how scientific papers publicly connect themselves to AI-related literature, but they do not fully capture the actual use of AI tools, models, code, or tacit expertise. Second, the identification of AI-related literature relies on the OpenAlex Artificial Intelligence subfield, but this taxonomy has several limitations. It may miss related papers or include papers only loosely related to AI. Third, our measure of institutional AI capability is based on CSRankings. This measure captures visible academic strength in AI research, but it may not fully represent applied AI capability, industry-oriented AI collaboration, and so on. Moreover, the analysis is observational. Although we control for paper-level, author-level, institutional, field, year, and document-type characteristics, the estimates should be interpreted as conditional associations rather than causal effects. Finally, several theoretical processes in this study are inferred rather than directly observed. Qualitative research could provide a clearer account of how AI knowledge is selected, adapted, and legitimized.

## References

Agrawal, A., McHale, J., Oettl, A., 2024. Artificial intelligence and scientific discovery: a model of prioritized search. Research Policy 53, 104989. https://doi.org/10.1016/j.respol.2024.104989

Berger, E.D., 2020. CSRankings.

Bianchini, S., Di Girolamo, V., Ravet, J., Arranz, D., 2026. AI in science: When and where it makes a difference. Research Policy 55, 105478.

https://doi.org/10.1016/j.respol.2026.105478

Bianchini, S., Müller, M., Pelletier, P., 2022. Artificial intelligence in science: An emerging general method of invention. Research Policy 51, 104604. https://doi.org/10.1016/j.respol.2022.104604

Blume, S., 1993. Correlates of Creativity: Striking the Mother Lode In Science. The Importance of Age, Place, and Time. Paula E. Stephan and Sharon G. Levin. Oxford University Press, New York, 1992. xiv, 194 pp., illus. $29.95. Science 259, 107–108. https://doi.org/10.1126/science.259.5091.107.a

Bornmann, L., Daniel, H., 2008. What do citation counts measure? A review of studies on citing behavior. Journal of Documentation 64, 45–80. https://doi.org/10.1108/00220410810844150

Boschma, R., 2005. Proximity and Innovation: A Critical Assessment. Regional Studies 39, 61–74. https://doi.org/10.1080/0034340052000320887

Bresnahan, T.F., Trajtenberg, M., 1995. General purpose technologies 'Engines of growth'? Journal of Econometrics 65, 83–108. https://doi.org/10.1016/0304-4076(94)01598-T

Burt, R.S., 1992. Structural Holes. Harvard University Press. https://doi.org/10.4159/9780674029095

Carlile, P., 2003. Transferring, Translating and Transforming: An Integrative Framework.

Carlile, P.R., 2004. Transferring, Translating, and Transforming: An Integrative Framework for Managing Knowledge Across Boundaries. Organization Science 15, 555–568. https://doi.org/10.1287/orsc.1040.0094

Chai, S., Menon, A., 2019. Breakthrough recognition: Bias against novelty and competition for attention. Research Policy 48, 733–747. https://doi.org/10.1016/j.respol.2018.11.006

Cohen, W.M., Levinthal, D.A., 1990. Absorptive Capacity: A New Perspective on Learning and Innovation. Administrative Science Quarterly 35, 128. https://doi.org/10.2307/2393553

Cui, H., Lin, Y., Wu, L., Evans, J.A., 2026. Aging and the narrowing of scientific

innovation. Science 392, 588–591. https://doi.org/10.1126/science.ady8732

DiMaggio, P., Hargittai, E., 2001. From the “Digital Divide” to “Digital Inequality”: Studying Internet Use as Penetration Increases. Princeton University, Woodrow Wilson School of Public and International Affairs, Center for Arts and Cultural Policy Studies., Working Papers.

Douze, M., Guzhva, A., Deng, C., Johnson, J., Szilvasy, G., Mazaré, P.-E., Lomeli, M., Hosseini, L., Jégou, H., 2025. The Faiss library. https://doi.org/10.48550/arXiv.2401.08281

Dunn, O.J., 1961. Multiple Comparisons among Means. Journal of the American Statistical Association 56, 52–64. https://doi.org/10.1080/01621459.1961.10482090

Fleming, L., 2001. Recombinant Uncertainty in Technological Search. Management Science 47, 117–132. https://doi.org/10.1287/mnsc.47.1.117.10671

Fleming, L., Waguespack, D.M., 2007. Brokerage, Boundary Spanning, and Leadership in Open Innovation Communities. Organization Science 18, 165–180. https://doi.org/10.1287/orsc.1060.0242

Funk, R.J., Owen-Smith, J., 2017. A Dynamic Network Measure of Technological Change. Management Science 63, 791–817. https://doi.org/10.1287/mnsc.2015.2366

Gao, J., Wang, D., 2024. Quantifying the use and potential benefits of artificial intelligence in scientific research. Nat Hum Behav 8, 2281–2292. https://doi.org/10.1038/s41562-024-02020-5

Garfield, E., 1972. Citation Analysis as a Tool in Journal Evaluation. Science 178, 471–479. https://doi.org/10.1126/science.178.4060.471

Hargadon, A., Sutton, R.I., 1997. Technology brokering and innovation in a product development firm. Administrative science quarterly 716–749.

Lin, Z., Yin, Y., Liu, L., Wang, D., 2023. SciSciNet: A large-scale open data lake for the science of science research. Sci Data 10, 315. https://doi.org/10.1038/s41597-023-02198-9

Liu, Y., Xie, Y., Shen, X., Wu, D., 2025. Artificial intelligence use and scientific

innovation. Asso for Info Science & Tech asi.70043. https://doi.org/10.1002/asi.70043

Merton, R., 1973. The Normative Structure of Science.

Merton, R.K., 1968. The Matthew Effect in Science. Science 159, 56–63. https://doi.org/10.1126/science.159.3810.56

Milgrom, P., Roberts, J., 1990. The Economics of Modern Manufacturing: Technology, Strategy, and Organization. American Economic Review 80, 511–28.

Ministry of Education of the People’s Republic of China. 2018. Artificial Intelligence Innovation Action Plan for Institutions of Higher Education [高等学校人工智能创新行动计划]. Ministry of Education Document No. 3, Beijing.

National Science Foundation. 2023. National Artificial Intelligence Research Institutes. Program Solicitation NSF 23-610.

National Science Foundation. 2024. National Artificial Intelligence Research Resource Pilot.

Nooteboom, B., Van Haverbeke, W., Duysters, G., Gilsing, V., van den Oord, A., 2007. Optimal cognitive distance and absorptive capacity. Research Policy 36, 1016–1034. https://doi.org/10.1016/j.respol.2007.04.003

Packalen, M., Bhattacharya, J., 2019. Age and the Trying Out of New Ideas. Journal of Human Capital 13, 341–373. https://doi.org/10.1086/703160

Priem, J., Piwowar, H., Orr, R., 2022. OpenAlex: A fully-open index of scholarly works, authors, venues, institutions, and concepts. https://doi.org/10.48550/arXiv.2205.01833

Rosenkopf, L., Nerkar, A., 2001. Beyond local search: boundary-spanning, exploration, and impact in the optical disk industry. Strategic Management Journal 22, 287–306. https://doi.org/10.1002/smj.160

Salatino, A.A., Thanapalasingam, T., Mannocci, A., Birukou, A., Osborne, F., Motta, E., 2020. The Computer Science Ontology: A Comprehensive Automatically-Generated Taxonomy of Research Areas. Data Intelligence 2, 379–416. https://doi.org/10.1162/dint_a_00055

Santos Silva, J.M.C., Tenreyro, S., 2011. Further simulation evidence on the performance of the Poisson pseudo-maximum likelihood estimator. Economics Letters 112, 220–222. https://doi.org/10.1016/j.econlet.2011.05.008

Schumpeter, J.A., 1983. The Theory of Economic Development: An Inquiry Into Profits, Capital, Credit, Interest, and the Business Cycle. Transaction Publishers.

Silva, J.M.C.S., Tenreyro, S., 2006. The Log of Gravity. The Review of Economics and Statistics 88, 641–658. https://doi.org/10.1162/rest.88.4.641

Singh, A., D'Arcy, M., Cohan, A., Downey, D., Feldman, S., 2023. SciRepEval: A Multi-Format Benchmark for Scientific Document Representations. https://doi.org/10.48550/arXiv.2211.13308

State Council of the People's Republic of China. 2017. New Generation Artificial Intelligence Development Plan [新一代人工智能发展规划]. State Council Document No. 35, Beijing.

Suchman, M.C., 1995. Managing Legitimacy: Strategic and Institutional Approaches. The Academy of Management Review 20, 571–610. https://doi.org/10.2307/258788

Teece, D.J., 1986. Profiting from technological innovation: Implications for integration, collaboration, licensing and public policy. Research Policy 15, 285–305. https://doi.org/10.1016/0048-7333(86)90027-2

Tushman, M.L., 1977. Special Boundary Roles in the Innovation Process. Administrative Science Quarterly 22, 587. https://doi.org/10.2307/2392402

U.S. Congress. 2021. National Artificial Intelligence Initiative Act of 2020. Pub. L. No. 116–283, Division E.

Uzzi, B., Mukherjee, S., Stringer, M., Jones, B., 2013. Atypical Combinations and Scientific Impact. Science 342, 468–472. https://doi.org/10.1126/science.1240474

van Deursen, A.J.A.M., Helsper, E.J., 2015. The Third-Level Digital Divide: Who Benefits Most from Being Online?, in: Communication and Information Technologies Annual. Emerald Group Publishing Limited, p. 0. https://doi.org/10.1108/S2050-206020150000010002

Wang, H., Fu, T., Du, Y., Gao, W., Huang, K., Liu, Z., Chandak, P., Liu, S., Van Katwyk, P., Deac, A., Anandkumar, A., Bergen, K., Gomes, C.P., Ho, S., Kohli, P., Lasenby, J., Leskovec, J., Liu, T.-Y., Manrai, A., Marks, D., Ramsundar, B., Song, L., Sun, J., Tang, J., Veličković, P., Welling, M., Zhang, L., Coley, C.W., Bengio, Y., Zitnik, M., 2023. Scientific discovery in the age of artificial intelligence. Nature 620, 47–60. https://doi.org/10.1038/s41586-023-06221-2

Welch, B.L., 1947. The generalisation of student's problems when several different population variances are involved. Biometrika 34, 28–35. https://doi.org/10.1093/biomet/34.1-2.28

Whitley, R., 2000. The Intellectual and Social Organization of the Sciences (Second Edition: with new introductory chapter entitled 'Science Transformed? The Changing Nature of Knowledge Production at the End of the Twentieth Century').

Wilinski, J., 2026. The Science Data Lake: A Unified Open Infrastructure Integrating 293 Million Papers Across Eight Scholarly Sources with Embedding-Based Ontology Alignment. https://doi.org/10.48550/arXiv.2603.03126

Wu, L., Wang, D., Evans, J.A., 2019. Large teams develop and small teams disrupt science and technology. Nature 566, 378–382. https://doi.org/10.1038/s41586-019-0941-9

Wu, L., Yi, F., Bu, Y., Lu, W., Huang, Y., 2024. Toward scientific collaboration: A cost-benefit perspective. Research Policy 53, 104943. https://doi.org/10.1016/j.respol.2023.104943

Zhao, N., Li, L., 2026. The diverging effect of prestige and experience on the use of artificial intelligence knowledge. Asso for Info Science & Tech 77, 580–595. https://doi.org/10.1002/asi.70068

Zuckerman, H., Merton, R.K., 1972. Age, aging, and age structure in science. Higher Education 4, 1–4.

## Supplementary Information

*Supplementary Tables.*

### Table S1 OpenAlex Topics Included in the Artificial Intelligence Subfield

| No. | OpenAlex topic | No. | OpenAlex topic |
|---|---|---|---|
| 1 | Adaptation to Concept Drift in Data Streams | 40 | Imbalanced Data Classification Techniques |
| 2 | Advanced Clustering Algorithms Research | 41 | Intelligent Tutoring Systems and Adaptive Learning |
| 3 | Advanced Computational Techniques and Applications | 42 | Internet Traffic Analysis and Secure E-voting |
| 4 | Advanced Graph Neural Networks | 43 | Intuitionistic Fuzzy Systems Applications |
| 5 | Advanced Software Engineering Methodologies | 44 | Law, AI, and Intellectual Property |
| 6 | Advanced Technologies in Various Fields | 45 | Linguistic Studies and Language Acquisition |
| 7 | Advanced Text Analysis Techniques | 46 | Logic, programming, and type systems |
| 8 | Adversarial Robustness in Machine Learning | 47 | Logic, Reasoning, and Knowledge |
| 9 | AI and Multimedia in Education | 48 | Machine Learning and Algorithms |
| 10 | AI in cancer detection | 49 | Machine Learning and Data Classification |
| 11 | AI in Service Interactions | 50 | Machine Learning and ELM |
| 12 | AI-based Problem Solving and Planning | 51 | Machine Learning in Healthcare |
| 13 | Algorithms and Data Compression | 52 | Mathematical Control Systems and Analysis |
| 14 | Anomaly Detection Techniques and Applications | 53 | Metaheuristic Optimization Algorithms Research |
| 15 | Artificial Intelligence Applications | 54 | Multi-Agent Systems and Negotiation |
| 16 | Artificial Intelligence in Games | 55 | Natural Language Processing Techniques |
| 17 | Authorship Attribution and Profiling | 56 | Neural Networks and Applications |
| 18 | Bayesian Methods and Mixture Models | 57 | Neural Networks and Reservoir Computing |
| 19 | Bayesian Modeling and Causal Inference | 58 | Organizational and Employee Performance |
| 20 | Coding theory and cryptography | 59 | Privacy-Preserving Technologies in Data |
| 21 | Cognitive Computing and Networks | 60 | Psychiatry, Mental Health, Neuroscience |
| 22 | Cognitive Science and Mapping | 61 | Quantum Computing Algorithms and Architecture |
| 23 | Computational Physics and Python Applications | 62 | Quantum Information and Cryptography |
| 24 | Computer Science and Engineering | 63 | Reinforcement Learning in Robotics |
| 25 | Cryptographic Implementations and Security | 64 | Security and Verification in Computing |
| 26 | Cryptography and Data Security | 65 | Seismology and Earthquake Studies |
| 27 | Data Analysis with R | 66 | Semantic Web and Ontologies |
| 28 | Diverse Interdisciplinary Research Studies | 67 | Sentiment Analysis and Opinion Mining |
| 29 | Domain Adaptation and Few-Shot Learning | 68 | Solar Radiation and Photovoltaics |
| 30 | Edcuational Technology Systems | 69 | Speech and dialogue systems |
| 31 | Educational Robotics and Engineering | 70 | Speech Recognition and Synthesis |
| 32 | Educational Technology and Pedagogy | 71 | Statistical and Computational Modeling |
| 33 | Evolutionary Algorithms and Applications | 72 | Stochastic Gradient Optimization Techniques |
| 34 | Experience-Based Knowledge Management | 73 | Target Tracking and Data Fusion in Sensor Networks |
| 35 | Explainable Artificial Intelligence (XAI) | 74 | Text and Document Classification Technologies |
| 36 | Fuzzy Logic and Control Systems | 75 | Text Readability and Simplification |
| 37 | Gaussian Processes and Bayesian Inference | 76 | Topic Modeling |
| 38 | Geochemistry and Geologic Mapping | 77 | Wireless Signal Modulation Classification |
| 39 | Hate Speech and Cyberbullying Detection | | |

*Notes:* This table lists the 77 OpenAlex topics used to identify AI-related references.

**Table S2 Summary Statistics**

| | N | Mean | SD | Median | Max |
|---|---|---|---|---|---|
| C5 | 12827989 | 18.011 | 52.011 | 7.000 | 27642.000 |
| ln(C5+1) | 12827989 | 2.033 | 1.338 | 2.079 | 5.088 |
| AIRef | 12827989 | 0.093 | 0.290 | 0.000 | 1.000 |
| AI score | 12827989 | 0.008 | 0.047 | 0.000 | 1.000 |
| AI score$^2$ | 12827989 | 0.002 | 0.031 | 0.000 | 1.000 |
| ln(total refs+1) | 12827989 | 3.322 | 0.962 | 3.497 | 5.220 |
| Team size | 12827989 | 4.058 | 3.493 | 3.000 | 18.000 |
| h-index | 12827989 | 23.844 | 24.168 | 17.000 | 434.000 |
| Institution count | 12827989 | 2.111 | 1.927 | 2.000 | 11.000 |

*Notes:* This table reports the summary statistics for some variables.

**Table S3 Correlation matrix**

| | C5 | ln(C5+1) | AIRef | AI score | AI score^2 | ln(refs count+1) | Team size | h-index | Institution count |
|---|---|---|---|---|---|---|---|---|---|
| C5 | 1.000 | | | | | | | | |
| ln(C5+1) | 0.454*** | 1.000 | | | | | | | |
| AIRef | 0.007*** | 0.011*** | 1.000 | | | | | | |
| AI score | -0.012*** | -0.033*** | 0.495*** | 1.000 | | | | | |
| AI score^2 | -0.008*** | -0.025*** | 0.235*** | 0.899*** | 1.000 | | | | |
| ln(refs count+1) | 0.168*** | 0.413*** | 0.131*** | -0.027*** | -0.039*** | 1.000 | | | |
| Team size | 0.162*** | 0.423*** | -0.032*** | -0.048*** | -0.031*** | 0.231*** | 1.000 | | |
| h-index | 0.157*** | 0.215*** | -0.017*** | -0.014*** | -0.007*** | 0.042*** | 0.068*** | 1.000 | |
| Institution count | 0.147*** | 0.373*** | 0.022*** | -0.023*** | -0.019*** | 0.231*** | 0.692*** | 0.142*** | 1.000 |

*Notes.* This table reports pairwise Pearson correlation coefficients among the variables. * $p < 0.10$, ** $p < 0.05$, *** $p < 0.01$

**Table S4 The Impact of AI References on Citations**

| | (1) | (2) | (3) |
|---|---|---|---|
| | ln(C5+1) | ln(C5+1) | ln(C5+1) |
| AIRef | 0.010*** | | -0.009*** |
| | (0.004) | | (0.003) |
| AI score | | 0.217*** | 0.243*** |
| | | (0.021) | (0.019) |
| h-index | 0.006*** | 0.006*** | 0.006*** |
| | (0.000) | (0.000) | (0.000) |
| ln(refs count+1) | 0.478*** | 0.479*** | 0.479*** |
| | (0.005) | (0.005) | (0.005) |
| Team size | 0.089*** | 0.089*** | 0.089*** |
| | (0.001) | (0.001) | (0.001) |
| Institution count | 0.046*** | 0.046*** | 0.046*** |
| | (0.002) | (0.002) | (0.002) |
| Constant | -0.163*** | -0.166*** | -0.166*** |
| | (0.020) | (0.020) | (0.020) |
| Observations | 12827989 | 12827989 | 12827989 |
| $R^2$ | 0.497 | 0.497 | 0.497 |

*Notes:* 1. Clustered standard errors at the institution level are reported in parentheses; 2. * $p < 0.1$, ** $p < 0.05$, *** $p < 0.01$

**Table S5 PPML Regression by institutional AI capability group**

| | (1) | (2) | (3) |
|---|---|---|---|
| | T1 Low | T2 Middle | T3 High |
| AIRef | 0.003 | $0.033^{**}$ | $0.046^{***}$ |
| | (0.007) | (0.013) | (0.015) |
| AI score | $0.534^{***}$ | $0.627^{***}$ | $0.485^{***}$ |
| | (0.065) | (0.060) | (0.053) |
| Age | $-0.011^{***}$ | $-0.010^{***}$ | $-0.013^{***}$ |
| | (0.001) | (0.001) | (0.001) |
| $Age^2$ | $0.000^{***}$ | $0.000^{***}$ | $0.000^{***}$ |
| | (0.000) | (0.000) | (0.000) |
| ln(refs count+1) | $0.596^{***}$ | $0.585^{***}$ | $0.618^{***}$ |
| | (0.009) | (0.015) | (0.020) |
| Team size | $0.066^{***}$ | $0.073^{***}$ | $0.079^{***}$ |
| | (0.002) | (0.002) | (0.002) |
| h-index | $0.013^{***}$ | $0.013^{***}$ | $0.013^{***}$ |
| | (0.000) | (0.000) | (0.000) |
| Institution count | $0.020^{***}$ | $0.014^{***}$ | 0.001 |
| | (0.003) | (0.003) | (0.005) |
| Observations | 4,007,881 | 3,984,399 | 3,950,056 |
| Pseudo $R^2$ | 0.3706 | 0.3654 | 0.3502 |

*Notes:* 1. Clustered standard errors at the institution level are reported in parentheses; 2. $^{*}$ $p < 0.1$, $^{**}$ $p < 0.05$, $^{***}$ $p < 0.01$